\documentclass[aps,pra,reprint,showpacs,superscriptaddress]{revtex4-1}
\usepackage[english]{babel}
\usepackage[utf8]{inputenc}
\usepackage{amsthm}
\usepackage{xcolor}
\usepackage{mathtools}
\usepackage{graphicx}
\usepackage[T1]{fontenc}
\usepackage{siunitx}
\usepackage{natbib}
\usepackage{hyperref}
\hypersetup{
	colorlinks   = true, 
	urlcolor     = blue, 
	linkcolor    = blue,
	citecolor   = blue 
}

\begin{document}

\title{Optically levitated rotor at its thermal limit of frequency stability}

\author{Fons van der Laan}
\email[Corresponding author: ]{vfons@ethz.ch}
\author{Ren{\'e} Reimann}
\author{Andrei Militaru}
\author{Felix Tebbenjohanns}
\author{Dominik Windey}
\author{Martin Frimmer}
\author{Lukas Novotny}
\affiliation{Photonics Laboratory, ETH Z{\"u}rich, 8093 Zurich, Switzerland}

\date{6 July 2020}

\begin{abstract}
Optically levitated rotors are prime candidates for torque sensors whose precision is limited by the fluctuations of the rotation frequency. In this work, we investigate an optically levitated rotor at its fundamental thermal limit of frequency stability, where rotation-frequency fluctuations arise solely due to coupling to the thermal bath.
\end{abstract}

\maketitle

\section{Introduction} \label{sec:introduction}

Optically trapped particles have emerged as a versatile platform to study mechanical degrees of freedom driven by fluctuating forces arising from the coupling to a thermal bath~\cite{Li2011, Gieseler2014, Dechant2015, Ricci2017, Rondin2017, Gieseler2018}. Thus far, the levitodynamics community has mostly focused on the center-of-mass (COM) degrees of freedom~\cite{Chang2010,Romero-Isart2011,Gieseler2012,Kiesel2013,Millen2015,Jain2016,Vovrosh2017,Monteiro2017, Windey2019,Delic2019,Tebbenjohanns2020}. More recently, the rotational~\cite{Arita2013, Arita2015, Hoang2016, Kuhn2015, Kuhn2017control, Kuhn2017stable, Reimann2018, Ahn2018, Monteiro2018, Rider2019, Delord2019, Ahn2020} degrees of freedom also have moved to the center of attention.

In a linearly polarized light field, an anisotropic scatterer aligns to the polarization direction~\cite{Hoang2016}. The optical torque thus corresponds to a restoring force which, to first order, is linear in orientation angle, making this libration degree of freedom a harmonic oscillator~\cite{Delord2019}. In stark contrast, in a circularly polarized field, an anistropic scatterer experiences an orientation-independent torque, which sets the particle into continuous rotation. The dynamics of such a free rotor is distinctly different from the harmonic-oscillator physics of the COM or libration degree of freedom. Besides their promise to allow for the investigation of fundamental effects~\cite{Shi2016,Stickler2018,Stickler2018rotationquantum}, freely rotating nanoparticles in optical traps have recently attracted considerable attention  as the fastest rotating man-made objects~\cite{Reimann2018, Ahn2018, Ahn2020} and have been identified as potential candidates for pressure~\cite{Kuhn2017stable, Blakemore2020}, acceleration, and various torque-sensing schemes~\cite{Manjavacas2010, Zhao2012, Manjavacas2017, Xu2017, Ahn2020}. On the one hand, phase-locked driven rotors have been considered for torque sensing~\cite{Kuhn2017stable, Blakemore2020}, but the sensitivity of this scheme remains largely unexplored. On the other hand, the current state-of-the-art levitated torque-sensing technique detects changes in rotation frequency, such that its sensitivity depends on the stability of that frequency~\cite{Ahn2020}. At the current stage (where thermal forces dominate over measurement backaction), thermal fluctuations generated by the bath are expected to limit torque-sensing schemes based on optical levitation. Surprisingly, a study of the thermal fluctuations of an optically driven rotor has not been performed to date.

In this work, we experimentally investigate the fluctuations of the rotation frequency of an optically levitated nanorotor and provide an avenue for operating at the thermal limit of frequency stability. We find that reaching this limit at high rotation frequencies requires COM cooling. We establish that our system operates at the thermal limit of frequency stability by confirming that the frequency fluctuations scale in accordance with the fluctuation-dissipation theorem.

\section{Experimental Setup.} \label{sec:setup}
\begin{figure}
    \centering
    \includegraphics[width = \linewidth]{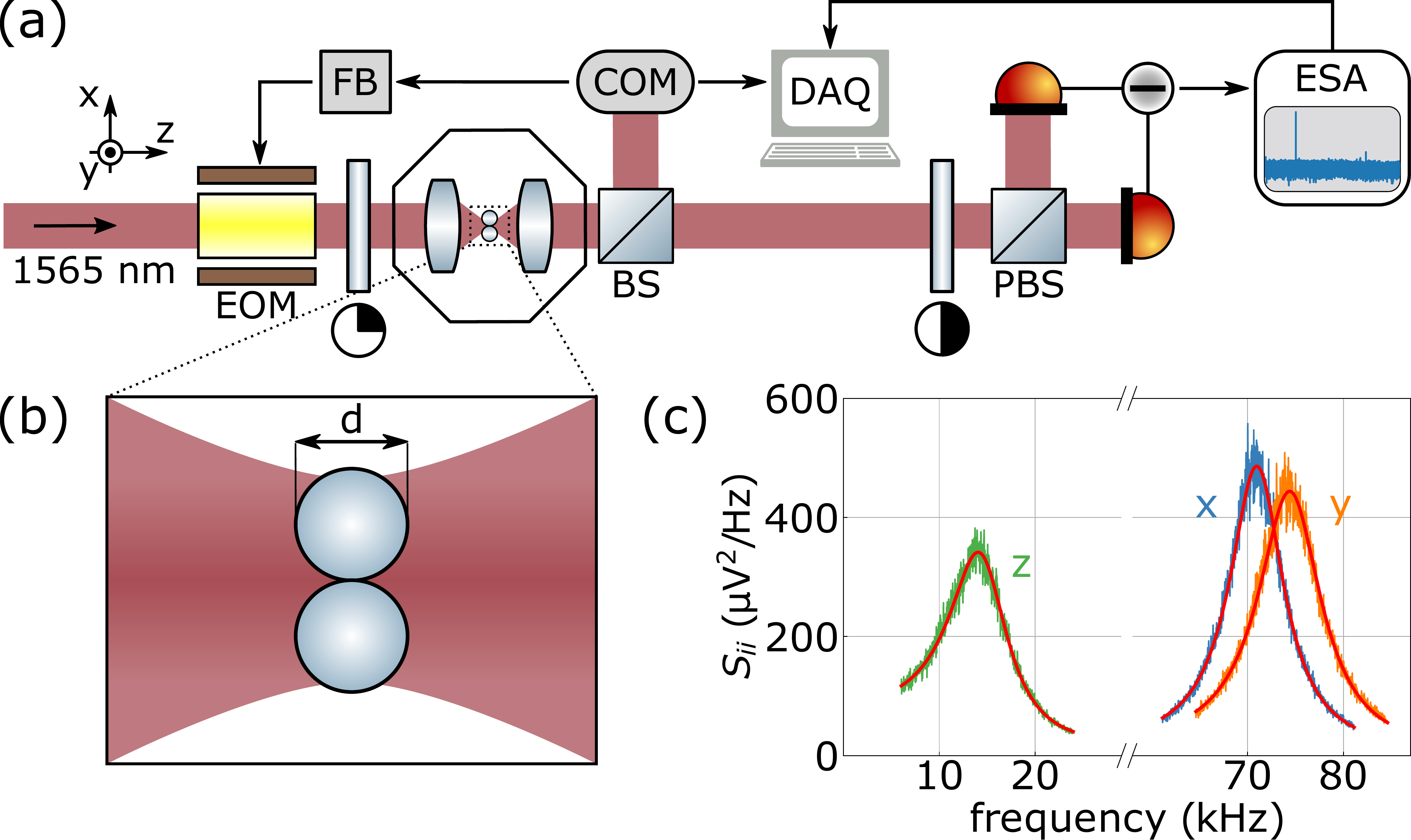}
    \caption{(a)~Simplified schematic of the experimental setup. Inside a vacuum chamber, an optical trap is formed by focusing a laser beam with an aspheric lens ($\mathrm{NA}=\SI{0.77}{}$). The intensity of the laser beam [wavelength $\lambda = \SI{1565.0(1)}{\nm}$] can be modulated with an electro-optical modulator (EOM). The polarization of the laser beam is set with a quarter-wave plate. The light is collected and split into two paths with a non-polarizing beamsplitter (BS). One half of the optical power is sent to a center-of-mass (COM) motion detector. The other half is used to measure the rotation in a balanced detection scheme. (b)~In the focus, we trap a dumbbell formed by two spherical nanoparticles with diameter $d$. For a circularly polarized trapping beam and at low enough pressure, the particle rotates due to the optical torque exerted by the laser beam~\cite{Kuhn2017control}. (c)~Power spectral densities of the COM motion of a rotor at a pressure $p_\mathrm{gas} = \SI{7.0(7)}{\milli\bar}$. The trapping laser is close to linearly polarized ($x$ axis), which orients the particle along the $x$ axis. The ratio between the damping rates along the $x$ and $y$ axes is $\SI{1.20(5)}{}$, identifying the trapped object as a dumbbell~\cite{Ahn2018}.}
    \label{fig:setup}
\end{figure}

The experimental setup is shown in Fig.~\ref{fig:setup}(a). 
We trap a single nanorotor in a strongly focused laser beam in vacuum. The polarization of the laser can be tuned from linear to circular by a quarter-wave plate. The forward scattered light is collected and sent to a detector which records the COM motion of the particle in the focal plane (along the $x$ and $y$ axes), and along the optical axis ($z$). The power spectral densities $S_{ii}$ ($i \in \{x, y, z\}$) of the COM motion display a Lorentzian shape, as shown in Fig.~\ref{fig:setup}(c), which is a signature of a harmonic trapping potential~\cite{Hebestreit2018Calibration}. 
The rotors in our trap are dumbbells composed of two nominally identical spherical silica nanoparticles loaded into the trap from a dispersion using a nebulizer~\cite{GieselerThesis}. The concentration of the dispersion is chosen to maximize the probability of two particles per aerosol droplet.
We verify that the trapped rotor is a dumbbell by comparing the measured damping rates of the COM motion along the $x$ and $y$ directions, while the dumbbell's long axis is aligned to the $x$ axis~\cite{Ahn2018}. Particles with equal damping rates along the $x$ and $y$ directions are removed from the trap and are not used in this work. We use the acquired COM position signals to parametrically feedback cool (FB) the COM motion of all three axes by modulating the laser power with an electro-optical modulator (EOM)~\cite{Jain2016}. 
We detect the rotor's angular orientation by sending the light exiting the vacuum chamber through a polarizing beamsplitter (PBS) and onto a balanced photodetector (bandwidth \SI{1.6}{\giga\Hz})~\cite{Reimann2018}. Using a half-wave plate, we balance the photodetector signal and send it to an electronic spectrum analyzer (ESA, bandwidth \SI{26.5}{\giga\Hz}). The pressure $p_\mathrm{gas}$ in the vacuum chamber is monitored using a Pirani gauge. The chamber temperature $T$ can be controlled with heating pads and is monitored by a sensor inside the vacuum chamber. Unless stated otherwise, all measurements are performed at room temperature using dumbbells with nominal diameter $d = \SI{136}{\nano\meter}$, a focal power of $P = \SI{0.27(2)}{\W}$, at a pressure of $p_{\rm gas} = \SI{5.0(5)E-2}{\milli\bar}$, and with a circularly polarized trapping beam.

\section{Results} \label{sec:results}
\subsection{Role of COM cooling}
\begin{figure}
    \centering
    \includegraphics[width =\linewidth]{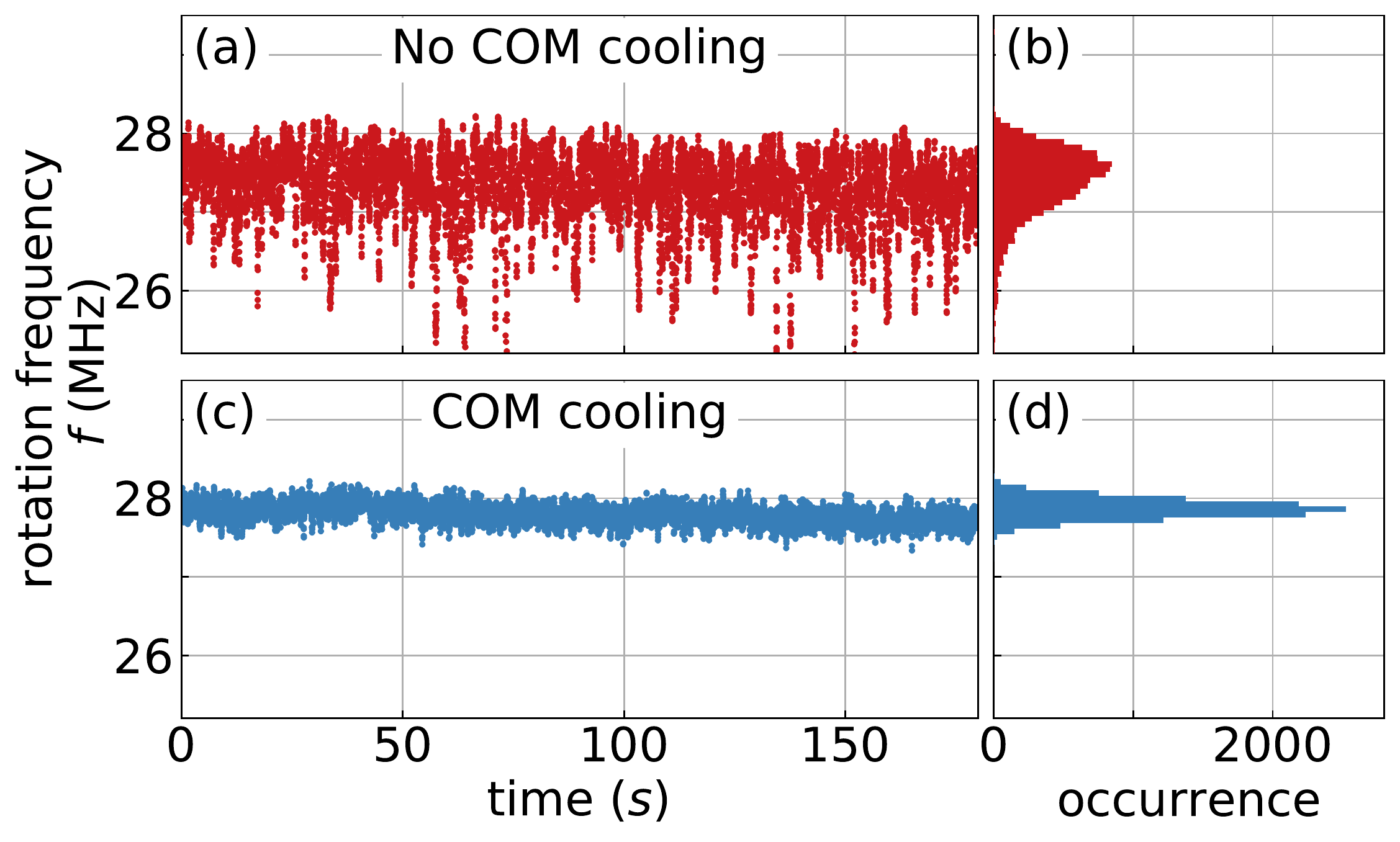}
    \caption{Effect of COM cooling on rotation frequency $f$. 
    (a)~Time trace of $f$ for a rotor at pressure $p_\mathrm{gas} = \SI{5.0(5)E-3}{\milli\bar}$ without COM cooling. 
    (b)~Histogram of time trace shown in~(a).
    (c)~Time trace at the same pressure as in (a)~but under COM cooling. 
    (d)~Histogram of time trace shown in~(c).
    The motion along $x$ and $y$ is cooled below $\SI{20}{\K}$, while the motion along $z$ is cooled to $\SI{90}{\K}$.}
    \label{fig:compare_cooling}
\end{figure}
We start by investigating the effect of COM cooling on the rotation frequency of a levitated dumbbell. For a rotation frequency $f$, the spectrum analyzer (ESA) shows a power spectral density with a narrow peak at $2f$~\cite{Reimann2018, Ahn2018}. 
In Fig.~\ref{fig:compare_cooling}(a), we show a measurement of $f$, extracted from the ESA spectrum, as a function of time at pressure $p_\mathrm{gas} = \SI{5.0(5)E-3}{\milli\bar}$ in the absence of COM feedback cooling. 
Figure~\ref{fig:compare_cooling}(b) shows a histogram of the frequency values of the time trace in Fig.~\ref{fig:compare_cooling}(a). We observe that the distribution is skewed towards smaller frequency values. This feature can be explained by the influence of the thermally driven COM motion. The dumbbell explores regions where the light intensity (and therefore the optical torque) is reduced as compared to the trap center. Accordingly, $f$ depends on the COM energy of the rotor.  To corroborate our conjecture, we show a measurement of the rotation frequency under COM feedback cooling in Figs.~\ref{fig:compare_cooling}(c)~and~\ref{fig:compare_cooling}(d). 
Indeed, the fluctuations of $f$ are strongly reduced by COM cooling and the distribution becomes symmetric. We conclude that the rotation frequency of an optically levitated dumbbell at room temperature can exhibit fluctuations arising from the thermal COM motion in the trapping potential.

\begin{figure*}
	\centering
	\includegraphics[width = \linewidth]{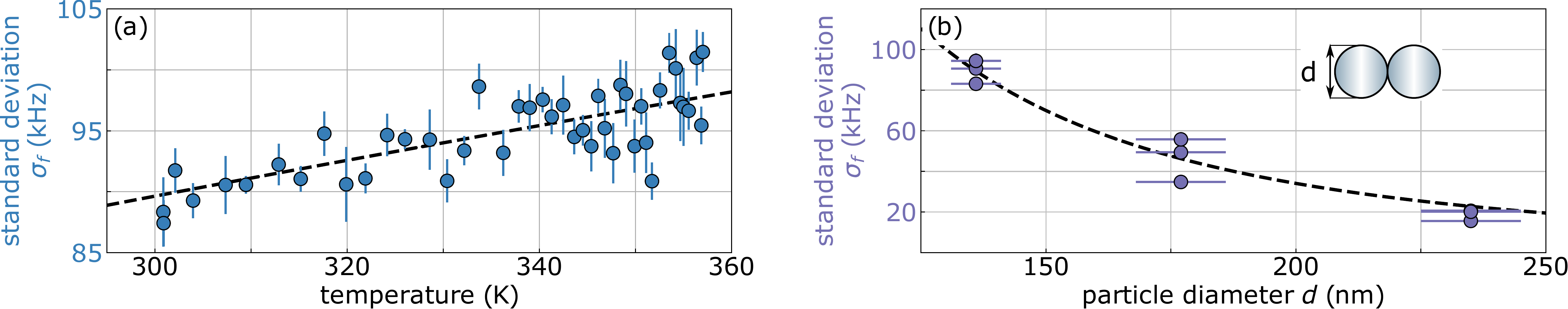}
	\caption{(a) Dependence of rotational fluctuations $\sigma_f$ on gas temperature $T$. Measurement of $\sigma_f$ as a function of $T$ at $p_\mathrm{gas} = \SI{9(1)E-2}{\milli\bar}$. The dashed line shows a fit to Eq.~\eqref{eq:std}. (b) Dependence of $\sigma_f$ on rotor size $d$. The dashed line shows the theoretical prediction according to Eq.~\eqref{eq:std} with a scaling factor extracted from (a) and using $T = \SI{300}{\K}$.}
	\label{fig:variance_temperature_size}
\end{figure*}
\subsection{Model}
In order to understand the rotation dynamics, we model the rotation frequency $f$ with the equation of motion
\begin{equation}
    2 \pi I \frac{\mathrm{d}}{\mathrm{d}t}f(t) + 2 \pi I \gamma f(t) = \tau_{\mathrm{opt}} + \tau_{\mathrm{th}}(t),
    \label{eq:EoM}
\end{equation}
where $t$ is the time, $I$ is the moment of inertia of the rotor, and $\gamma$ the rotational damping rate due to gas friction~\cite{Ahn2020}. A circularly polarized light field generates a constant optical torque $\tau_{\mathrm{opt}}$, which drives the rotation. Besides the optical torque, we include a fluctuating thermal torque $\tau_{\mathrm{th}}$ exerted by the surrounding gas. This stochastic thermal torque $\tau_\mathrm{th}$ has zero mean, is Gaussian distributed in magnitude, and is linked to $\gamma$ via the fluctuation-dissipation theorem
\begin{equation}
\langle \tau_{\mathrm{th}}(t)\tau_{\mathrm{th}}(t+t') \rangle_t = 2 I \gamma k_\mathrm{B} T \delta(t'),
\end{equation}
where $k_\mathrm{B}$ is the Boltzmann constant, $T$ is the temperature of the surrounding gas, and $\delta$ is the Dirac delta function~\cite{kubo1966fluctuation}. 
In steady state, the optical torque is balanced by the damping $\gamma$, resulting in a mean rotation frequency
\begin{equation}
    \langle f \rangle = \frac{1}{2 \pi} \frac{\tau_{\mathrm{opt}}}{I\,\gamma}.
    \label{eq:mean}
\end{equation}
The thermal torque $\tau_\mathrm{th}$ causes $f$ to fluctuate with standard deviation
\begin{equation}
\sigma_\mathrm{th}  = \frac{1}{2 \pi} \sqrt{\frac{k_{\mathrm{B}} T}{I}}.
\label{eq:std}
\end{equation}
Therefore, the thermal rotation-frequency fluctuations solely depend on the ratio of the temperature of the surrounding gas and the moment of inertia of the rotor. These fluctuations fundamentally limit the sensitivity of torque sensors using optically levitated rotors. In addition to thermal fluctuations, technically induced fluctuations $\sigma_\mathrm{tech}$ contribute to the measured rotation-frequency fluctuations according to $\sigma_f = \sqrt{\sigma_\mathrm{th}^2 + \sigma_\mathrm{tech}^2}$.
Importantly, $\sigma_f$ is only thermally limited if $\sigma_\mathrm{tech}$ can be neglected, i.e., $\sigma_\mathrm{tech} \ll \sigma_\mathrm{th}$. According to Eq.~\eqref{eq:mean}, technical fluctuations can arise from variations in the damping rate $\Delta \gamma$ and in the optical torque $\Delta \tau_\mathrm{opt}$, which yields
\begin{equation}
    \sigma_\mathrm{tech} = \sqrt{\left(\frac{\Delta \gamma}{\gamma} \langle f \rangle \right)^2 + \left( \frac{\Delta \tau_\mathrm{opt}}{\tau_\mathrm{opt}} \langle f \rangle \right)^2 }.
    \label{eq:std_tech}
\end{equation}
Importantly, technical noise contributions scale with the mean rotation frequency, such that it becomes increasingly difficult to operate at the fundamental thermal limit as the rotation frequency grows. This fact poses a severe challenge for measurement schemes requiring large rotation speeds, such as those aiming to detect vacuum friction~\cite{Manjavacas2017}. Mitigation strategies include careful pressure stabilization (to minimize $\Delta\gamma$), as well as active feedback cooling of the COM motion, in order to minimize $\Delta\tau_\text{opt}$. In the following, we show that we have suppressed technical contributions to frequency fluctuations and reached the thermal limit of frequency stability.

\subsection{Temperature dependence}
We start by measuring the standard deviation of the frequency fluctuations $\sigma_f$ as we change the temperature $T$ of the vacuum chamber [see Fig.~\ref{fig:variance_temperature_size}(a)]. 
We use a focal power of $P = \SI{0.57(2)}{\W}$ and keep $f$ below $\SI{3}{\mega\Hz}$ to limit technical contributions to the observed frequency fluctuations.
A detailed description of how we extract $\sigma_f$ throughout this work can be found in the Appendix. The standard deviation $\sigma_f$ increases for increasing temperature. We fit Eq.~\eqref{eq:std} to the data (dashed line) and extract the rotor's moment of inertia $I_{\rm exp} = \SI{1.31(1)E-32}{\kg \m \squared}$. For comparison, we calculate the moment of inertia of a nanodumbbell $I_{\rm theo} = (7/60) \pi \rho d^5$, where $\rho$ is the density of the particle material~\cite{Taylorbook}. Using the density of fused silica $\rho = \SI{2200}{\kg\per\m\cubed}$ and the nominal nanosphere diameter $d = \SI{136}{\nano\meter}$, we calculate $I_{\rm theo} = \SI{3.75E-32}{\kg \m \squared}$. Our theoretical result overestimates the moment of inertia. We can currently only speculate about the origin of this discrepancy. Possible explanations include dumbbells that (1) consist of spheres of different sizes, (2) have a finite contact area instead of a single contact point, or (3) experience a structural transition while being trapped~\cite{RicciThesis}.

\subsection{Size dependence}
To further test our understanding, we measure how the rotor's moment of inertia $I$ influences $\sigma_f$. To this end, we vary $I$ by using particles of different nominal diameters $d$. To ensure operation at the thermal limit for all rotors, we cool the COM motion.
Figure~\ref{fig:variance_temperature_size}(b) shows $\sigma_f$ for dumbbells consisting of spheres with nominal diameter $d$. As predicted by theory, $\sigma_f$ decreases with increasing particle diameter $d$. Each data point in Fig.~\ref{fig:variance_temperature_size}(b) corresponds to an individual dumbbell. We attribute the spread of $\sigma_f$ for one nominal diameter $d$ to particle size variations.
In Fig.~\ref{fig:variance_temperature_size}(b), we also include the prediction  of Eq.~\eqref{eq:std} (dashed line) with the moment of inertia corrected by the factor $I_{\rm exp}/I_{\rm theo}$ found in Fig.~\ref{fig:variance_temperature_size}(a). Our experimental data match the theory well. We stress that the dashed line does not rely on any free parameter. Remarkably, we observe that the correction factor $I_{\rm exp}/I_{\rm theo}$, extracted for a rotor with $d=\SI{136}{\nano\meter}$, applies to rotors of all measured sizes.

\subsection{Further checks}
\begin{figure}
    \centering
    \includegraphics[width =\linewidth]{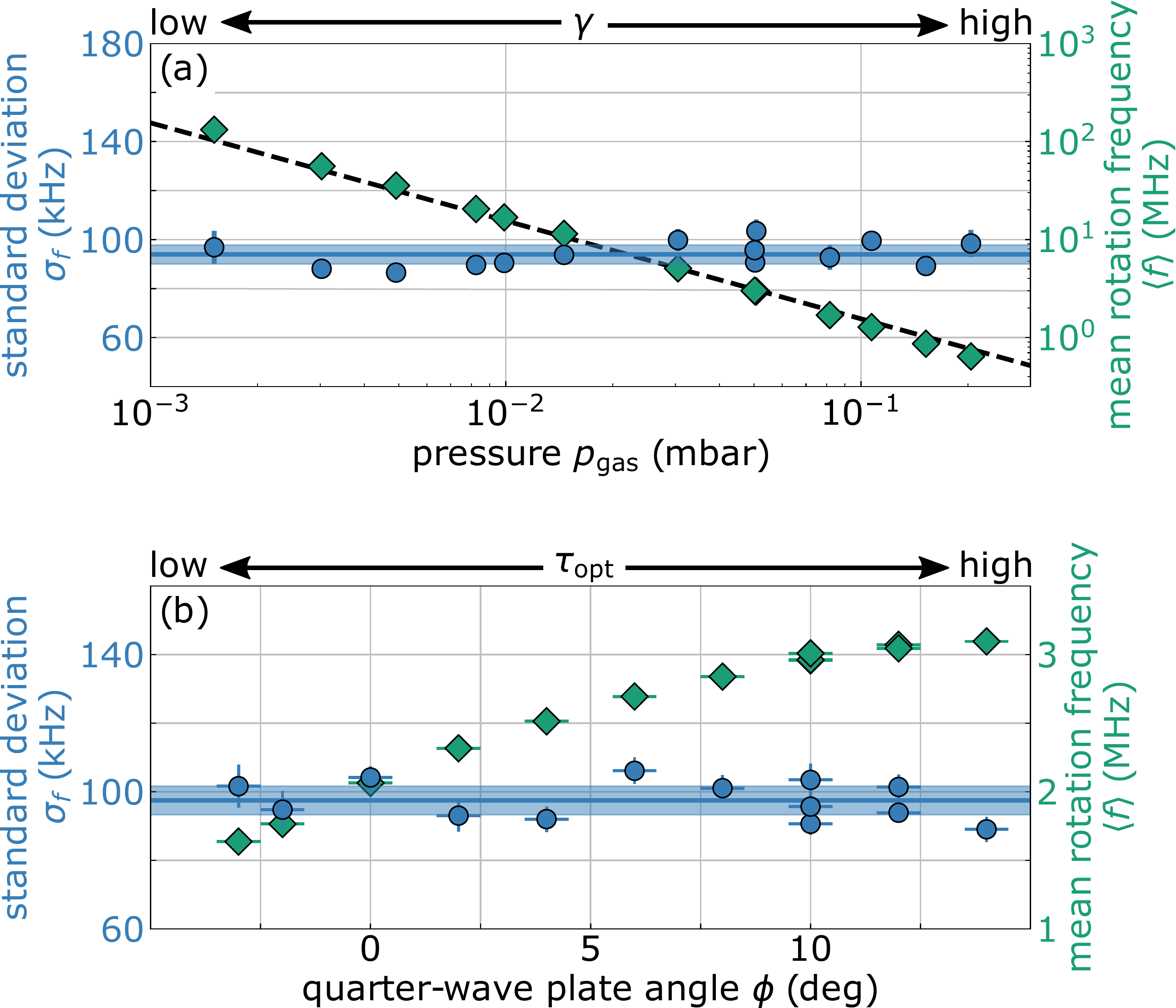}
    \caption{(a) Plot of the mean rotation frequency $\langle f \rangle$ (green diamonds) and its standard deviation $\sigma_f $(blue circles) as a function of gas pressure $p_\mathrm{gas}$. While $\sigma_f$ remains constant over a pressure range spanning more than two orders of magnitude, $\langle f \rangle$ is proportional to $1/p_\mathrm{gas}$ (shown as dashed line). (b) Influence of optical torque $\tau_\mathrm{opt}$. We measure $\langle f \rangle$ (green diamonds) and $\sigma_f$ (blue circles) as a function of quarter-wave plate angle $\phi$, setting the polarization state of the trapping field, and thus $\tau_{\rm opt}$ [cf. Fig.~\ref{fig:setup}(a)].}
    \label{fig:variance_pressure_waveplate}
\end{figure}

Having confirmed the scaling $\sigma_f\propto \sqrt{T/I}$, we turn our attention to the influence of the damping rate $\gamma$. According to Eq.~\eqref{eq:mean}, $\sigma_f$ does not depend on the damping rate $\gamma\propto p_{\rm gas}$ when $\sigma_f$ is thermally limited, and therefore neither on pressure $p_\mathrm{gas}$. Figure~\ref{fig:variance_pressure_waveplate}(a) shows the mean rotation frequency $f$ (green diamonds) and its standard deviation $\sigma_f$ (blue circles) as a function of $p_\mathrm{gas}$. We use COM feedback cooling to mitigate fluctuations due to variations in $\tau_\mathrm{opt}$. As we decrease $p_\mathrm{gas}$ from $\SI{2.0(2)E-1}{\milli\bar}$ to $\SI{1.5(2)E-3}{\milli\bar}$, the mean rotation frequency increases by two orders of magnitude and follows a $1/p_\mathrm{gas}$ dependence (fit shown as black dashed line)~\cite{Reimann2018, Ahn2018}. 
In stark contrast, and  as predicted by Eq.~\eqref{eq:std}, $\sigma_f$ does not depend on $p_\mathrm{gas}$. 
The mean of $\sigma_f$ observed at the different pressures is shown as a blue horizontal line with its uncertainty depicted by the shaded area. For values of $p_\mathrm{gas}$ below $\SI{2E-2}{\milli\bar}$, slow variations in pressure cause a slow drift in the rotation frequency $f$. A careful analysis of the power spectral density of the rotation frequency allows us to correct for these slow drifts, as detailed in the Appendix.

As a final check of operating at the thermal limit, we investigate the influence of the optical torque $\tau_\mathrm{opt}$ on $\sigma_f$. We set $\tau_\mathrm{opt}$ by tuning the polarization of the trapping laser via the angle $\phi$ of the quarter-wave plate before the optical trap. For $\phi = \SI{15}{\degree}$, the trap polarization is circular. In Fig.~\ref{fig:variance_pressure_waveplate}(b), we show $\langle f \rangle$ (green diamonds) and $\sigma_f$ (blue circles) as a function of $\phi$. For $\phi < \SI{-3}{\degree}$, the alignment torque (due to the linear polarization of the trapping beam) prevents full rotation such that the particle librates~\cite{Kuhn2017control}. As predicted by Eq.~\eqref{eq:std}, $\sigma_f$ remains constant for increasing values of $\phi$,  whereas, in accordance with Eq.~\eqref{eq:mean}, $\langle f \rangle$ increases.

\section{Conclusion} \label{sec:conclusion}
We have experimentally investigated the fluctuations of the rotation frequency of a nanorotor optically levitated in a circularly polarized laser field. 
Our results demonstrate that in the absence of center-of-mass cooling, the variation in the optical intensity experienced by the rotor due to its thermal oscillation amplitude gives rise to significant fluctuations in the rotation frequency. For high rotation frequencies, these technical fluctuations largely exceed the thermal fluctuations. This insight is of high relevance for torque-sensing schemes that rely on optically levitated rotors. Our work demonstrates that the thermal limit of torque sensing (as given by the fluctuation-dissipation theorem for the rotational degree of freedom) requires cooling of the center-of-mass motion in currently used optical levitation systems. 
Furthermore, we have investigated thermal fluctuations of the rotation frequency as a function of different system parameters. The standard deviation of these fluctuations shows a square-root scaling with temperature and moment of inertia. Finally, we have shown that the standard deviation of the thermal rotation-frequency fluctuations depends neither on pressure nor on optical torque (and thus not on the mean rotation frequency). 
In conclusion, we have demonstrated to operate our system at the thermal limit of rotation-frequency stability.

\begin{acknowledgments}
This research was supported by ERC-QMES (Grant No.\ 338763), the NCCR-QSIT program (Grant No.\ 51NF40-160591), and the European Union’s Horizon 2020 research and innovation programme under Grant No. 863132 (iQLev). We thank A.~Nardi for valuable discussions and S.~Papadopoulos for illustration support.
\end{acknowledgments}

\appendix

\section*{Appendix: Extraction of the standard deviation} \label{sec:methods}
In this appendix, we describe how we extract the standard deviation of the rotation frequency from our measurements.
We determine the standard deviation $\sigma_f$ of the rotation frequency $f$ with two distinct methods.
For both methods, we extract $\sigma_f$ from a time trace of the rotation frequency (measured with a sampling frequency of about $\SI{33}{\Hz}$)  by splitting the trace into $10$ segments of equal duration. We denote the standard deviation of segment $i$ with $\sigma_{f,i}$.
The mean of all $\sigma_{f,i}$ yields $\sigma_f$. The error of $\sigma_f$ is estimated by the standard deviation of the 10 values $\sigma_{f,i}$, divided by $\sqrt{10}$.

\begin{figure}[b]
    \centering
    \includegraphics[width = \linewidth]{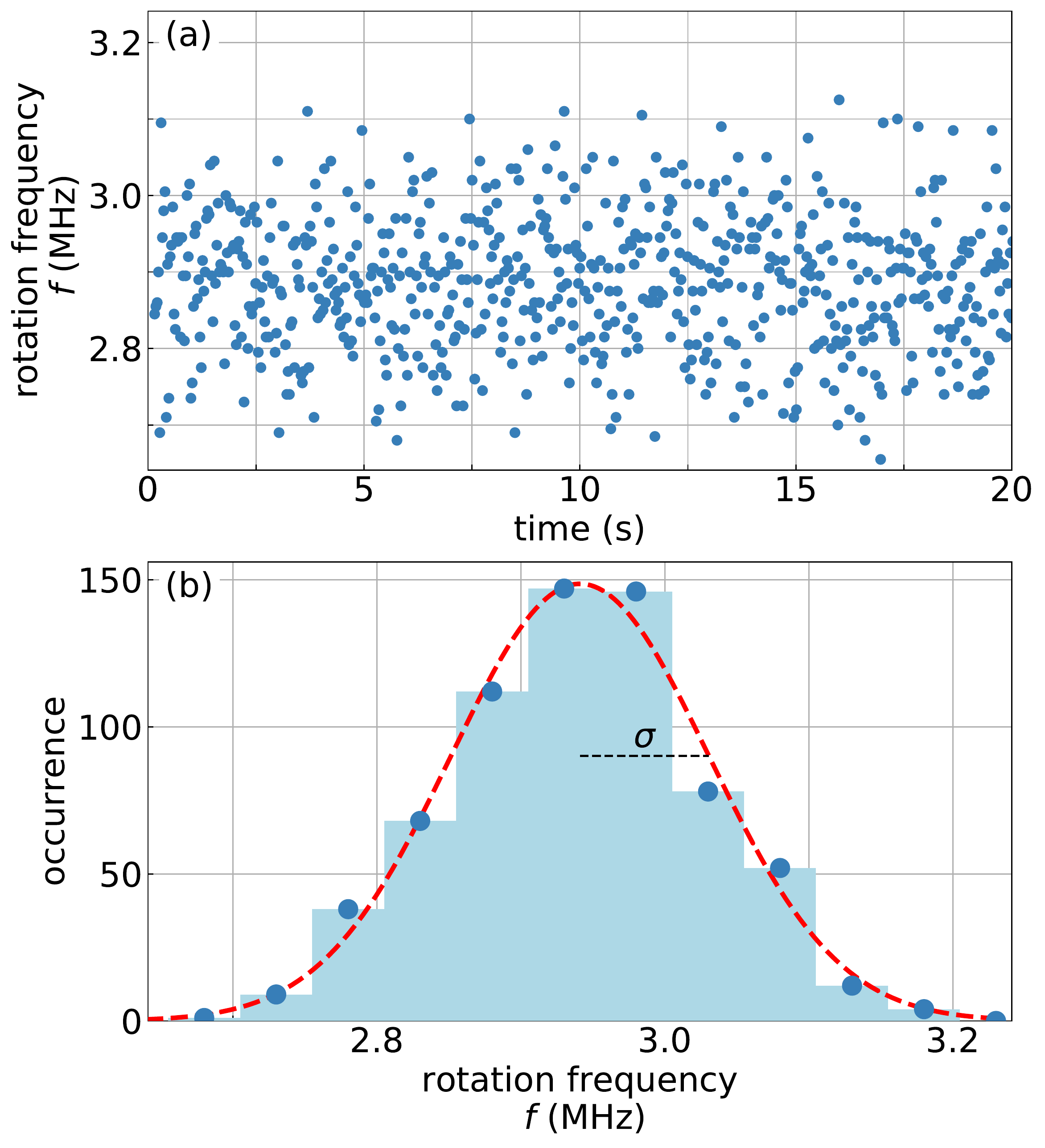}
    \caption{Method 1: Extraction of standard deviation $\sigma_{f,i}$ from a histogram of a time trace of the rotation frequency $f$. (a) Time trace of the rotation frequency $f$ of a dumbbell at pressure $p_\mathrm{gas} = \SI{5.0(5)E-2}{\milli\bar}$. (b) Histogram of time trace shown in (a). From a Gaussian fit (red dashed curve) we extract the standard deviation $\sigma_{f,i}$ of the time trace.}
    \label{fig:methods_hist}
\end{figure}

\subsection{Method 1} Method~1 extracts $\sigma_{f,i}$ from a histogram of the frequency values in time trace segment $i$.
As an example, we show a time trace of the rotation frequency $f$ in Fig.~\ref{fig:methods_hist}(a) together with the corresponding histogram of measured frequency values in Fig.~\ref{fig:methods_hist}(b). 
We fit this histogram with the Gaussian function
\begin{equation}
    h(f) = A e^{-\frac{(f - \langle f \rangle)^2}{2 {\sigma_{f,i}}^2}}
\end{equation}
and extract the amplitude $A$, the mean frequency $\langle f \rangle$, and the standard deviation $\sigma_{f,i}$ as free fit parameters.
Method 1 can be applied to data measured in the regime where the rotation-frequency fluctuations are thermally limited, i.e., $\sigma_\mathrm{th} \gg \sigma_\mathrm{tech}$. In this regime the histogram of the rotation frequency $f$ assumes a Gaussian shape. At lower pressures the technical fluctuations $\sigma_\mathrm{tech}$ become significant. As shown in Fig.~\ref{fig:methods_psd}(a), the mean rotation frequency $\langle f \rangle$ drifts more than $\sigma_\mathrm{th}$ due to slow pressure drifts. This drift broadens the corresponding histogram, depicted in Fig~\ref{fig:methods_psd}(b), to a non-Gaussian distribution. From the discussion in the main text, we understand that $\sigma_\mathrm{tech}$ depends linearly on mean rotation frequency $\langle f \rangle$ and thus inversely on pressure. Consequently, $\sigma_\mathrm{tech}$ becomes much larger than $\sigma_\mathrm{th}$ at low pressure, even though the relative drift of the damping rate $\Delta \gamma / \gamma$ is approximately constant for all pressures. Therefore method 1 fails in extracting $\sigma_{f,i}$ at low pressures.

\subsection{Method 2}
\begin{figure}
    \centering
    \includegraphics[width = \linewidth]{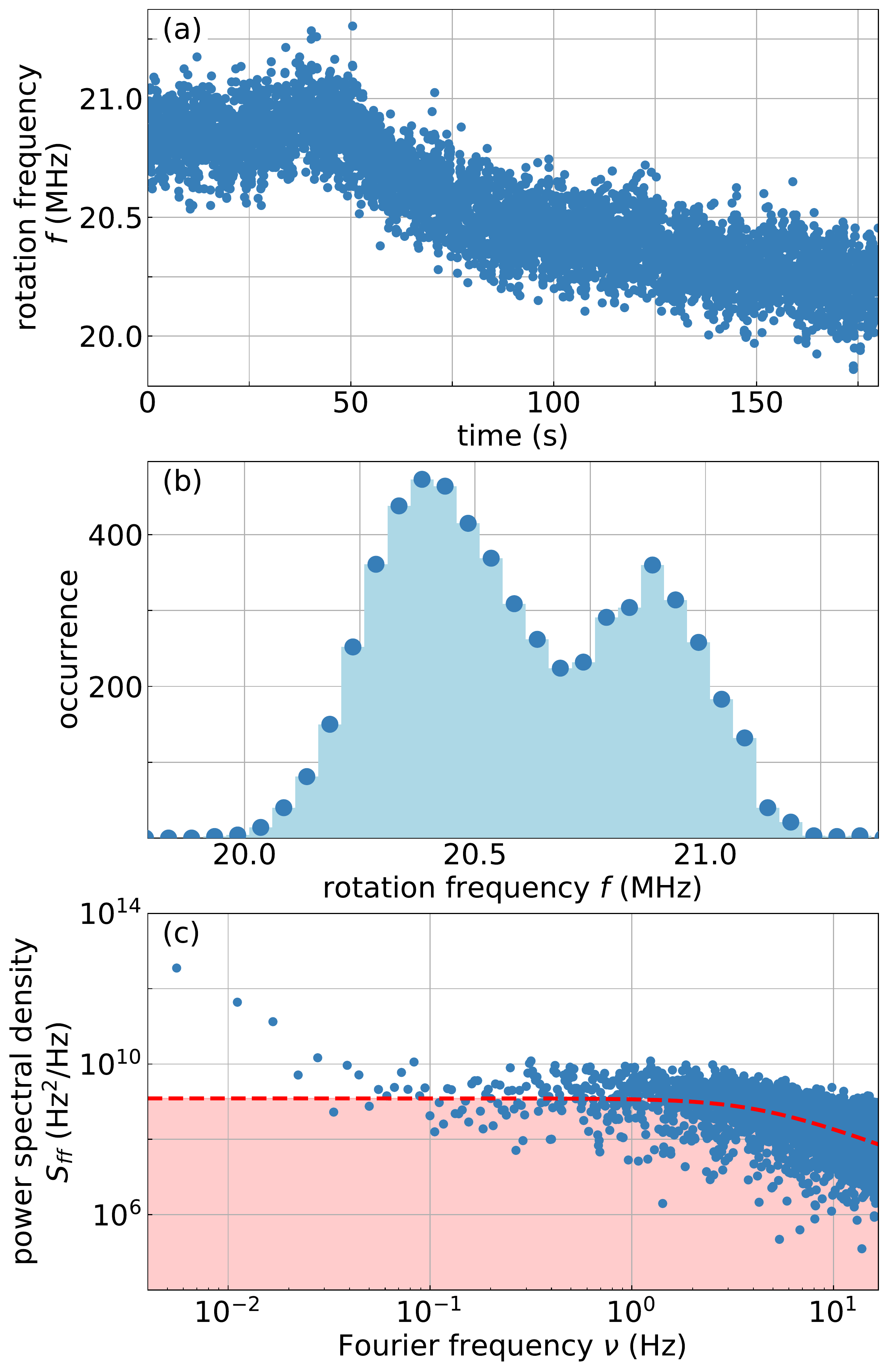}
    \caption{Method 2: Extraction of standard deviation $\sigma_f$ from $S_{ff}(\nu)$. (a) Time trace of the rotation frequency $f$ of a dumbbell at pressure $p_\mathrm{gas} = \SI{8.0(8)E-3}{\milli\bar}$. (b)  Histogram of time trace shown in (a). Because of slow pressure drifts, the histogram does not have a Gaussian shape. (c) Power spectral density $S_{ff}(\nu)$ of the time trace shown in (a). The standard deviation is extracted from the area under the Lorentzian fit (red dashed curve).}
    \label{fig:methods_psd}
\end{figure}
In this low-pressure regime, which for our system parameters starts at pressures smaller than $\SI{2.0(2)E-2}{\milli\bar}$, we apply method 2.
Method 2 is illustrated in Fig.~\ref{fig:methods_psd}(c) and utilizes the power spectral density $S_{ff}(\nu)$ of the time trace segment $i$ to determine $\sigma_{f,i}$.
From Eq.~\eqref{eq:EoM}, we find that $S_{ff}(\nu)$ has a Lorentzian shape for constant $\gamma$ and optical torque $\tau_\mathrm{opt}$. We therefore fit the Lorentzian function (red dashed line in Fig.~\ref{fig:methods_psd})
\begin{equation}
    h(\nu) = B \frac{\gamma}{\nu^2 + \gamma^2}
\end{equation}
to $S_{ff}(\nu)$, where amplitude $B$ and damping rate $\gamma$ are free fit parameters. 
Since the integral over the power spectral density of a signal is equal to the square of the standard deviation of this signal, we can integrate over the fitted Lorentzian to extract $\sigma_{f,i}$. By integrating over the fit (which ignores data in the low Fourier frequency regime, i.e., low $\nu$) we extract only thermal fluctuations and exclude effects from slow pressure drifts. Since the fit only deviates from the data in the low-$\nu$ regime, we conclude that $\sigma_f$ is thermally limited at Fourier frequencies  $\nu>0.1$~Hz. Method 2 suffers from two restrictions. First, the rotation frequency needs to be experimentally sampled at a rate larger than twice the damping rate $\gamma$ of the rotor in order to be able to resolve the Lorentzian shape. To understand this limitation, consider Fig.~\ref{fig:methods_psd}(c). With increasing pressure, the cut-off (which is at $
\gamma$) moves to higher frequency $\nu$ and will eventually fall out of the sampling window. Therefore, at high pressures, $\gamma$ becomes too large to be resolved and method 2 fails. 
Second, at very low pressures, $\gamma$ [and therefore the cut-off in $S_{ff}(\nu)$] moves to the low-$\nu$ regime, where $S_{ff}(\nu)$ is dominated by the effect of slow pressure drifts. Therefore, method~2 cannot be applied at too low pressures.

\bibliographystyle{apsrev4-1}
\bibliography{bibliography}

\begin{thebibliography}{42}%
\makeatletter
\providecommand \@ifxundefined [1]{%
 \@ifx{#1\undefined}
}%
\providecommand \@ifnum [1]{%
 \ifnum #1\expandafter \@firstoftwo
 \else \expandafter \@secondoftwo
 \fi
}%
\providecommand \@ifx [1]{%
 \ifx #1\expandafter \@firstoftwo
 \else \expandafter \@secondoftwo
 \fi
}%
\providecommand \natexlab [1]{#1}%
\providecommand \enquote  [1]{``#1''}%
\providecommand \bibnamefont  [1]{#1}%
\providecommand \bibfnamefont [1]{#1}%
\providecommand \citenamefont [1]{#1}%
\providecommand \href@noop [0]{\@secondoftwo}%
\providecommand \href [0]{\begingroup \@sanitize@url \@href}%
\providecommand \@href[1]{\@@startlink{#1}\@@href}%
\providecommand \@@href[1]{\endgroup#1\@@endlink}%
\providecommand \@sanitize@url [0]{\catcode `\\12\catcode `\$12\catcode
  `\&12\catcode `\#12\catcode `\^12\catcode `\_12\catcode `\%12\relax}%
\providecommand \@@startlink[1]{}%
\providecommand \@@endlink[0]{}%
\providecommand \url  [0]{\begingroup\@sanitize@url \@url }%
\providecommand \@url [1]{\endgroup\@href {#1}{\urlprefix }}%
\providecommand \urlprefix  [0]{URL }%
\providecommand \Eprint [0]{\href }%
\providecommand \doibase [0]{http://dx.doi.org/}%
\providecommand \selectlanguage [0]{\@gobble}%
\providecommand \bibinfo  [0]{\@secondoftwo}%
\providecommand \bibfield  [0]{\@secondoftwo}%
\providecommand \translation [1]{[#1]}%
\providecommand \BibitemOpen [0]{}%
\providecommand \bibitemStop [0]{}%
\providecommand \bibitemNoStop [0]{.\EOS\space}%
\providecommand \EOS [0]{\spacefactor3000\relax}%
\providecommand \BibitemShut  [1]{\csname bibitem#1\endcsname}%
\let\auto@bib@innerbib\@empty
\bibitem [{\citenamefont {Li}\ \emph {et~al.}(2011)\citenamefont {Li},
  \citenamefont {Kheifets},\ and\ \citenamefont {Raizen}}]{Li2011}%
  \BibitemOpen
  \bibfield  {author} {\bibinfo {author} {\bibfnamefont {T.}~\bibnamefont
  {Li}}, \bibinfo {author} {\bibfnamefont {S.}~\bibnamefont {Kheifets}}, \ and\
  \bibinfo {author} {\bibfnamefont {M.~G.}\ \bibnamefont {Raizen}},\ }\href
  {\doibase 10.1038/nphys1952} {\bibfield  {journal} {\bibinfo  {journal} {Nat.
  Phys.}\ }\textbf {\bibinfo {volume} {7}},\ \bibinfo {pages} {527} (\bibinfo
  {year} {2011})}\BibitemShut {NoStop}%
\bibitem [{\citenamefont {Gieseler}\ \emph {et~al.}(2014)\citenamefont
  {Gieseler}, \citenamefont {Quidant}, \citenamefont {Dellago},\ and\
  \citenamefont {Novotny}}]{Gieseler2014}%
  \BibitemOpen
  \bibfield  {author} {\bibinfo {author} {\bibfnamefont {J.}~\bibnamefont
  {Gieseler}}, \bibinfo {author} {\bibfnamefont {R.}~\bibnamefont {Quidant}},
  \bibinfo {author} {\bibfnamefont {C.}~\bibnamefont {Dellago}}, \ and\
  \bibinfo {author} {\bibfnamefont {L.}~\bibnamefont {Novotny}},\ }\href
  {\doibase 10.1038/nnano.2014.40} {\bibfield  {journal} {\bibinfo  {journal}
  {Nat. Nanotechnol.}\ }\textbf {\bibinfo {volume} {9}},\ \bibinfo {pages}
  {358} (\bibinfo {year} {2014})}\BibitemShut {NoStop}%
\bibitem [{\citenamefont {Dechant}\ \emph {et~al.}(2015)\citenamefont
  {Dechant}, \citenamefont {Kiesel},\ and\ \citenamefont {Lutz}}]{Dechant2015}%
  \BibitemOpen
  \bibfield  {author} {\bibinfo {author} {\bibfnamefont {A.}~\bibnamefont
  {Dechant}}, \bibinfo {author} {\bibfnamefont {N.}~\bibnamefont {Kiesel}}, \
  and\ \bibinfo {author} {\bibfnamefont {E.}~\bibnamefont {Lutz}},\ }\href
  {\doibase 10.1103/PhysRevLett.114.183602} {\bibfield  {journal} {\bibinfo
  {journal} {Phys. Rev. Lett.}\ }\textbf {\bibinfo {volume} {114}},\ \bibinfo
  {pages} {183602} (\bibinfo {year} {2015})}\BibitemShut {NoStop}%
\bibitem [{\citenamefont {Ricci}\ \emph {et~al.}(2017)\citenamefont {Ricci},
  \citenamefont {Rica}, \citenamefont {Spasenovi{\'{c}}}, \citenamefont
  {Gieseler}, \citenamefont {Rondin}, \citenamefont {Novotny},\ and\
  \citenamefont {Quidant}}]{Ricci2017}%
  \BibitemOpen
  \bibfield  {author} {\bibinfo {author} {\bibfnamefont {F.}~\bibnamefont
  {Ricci}}, \bibinfo {author} {\bibfnamefont {R.~A.}\ \bibnamefont {Rica}},
  \bibinfo {author} {\bibfnamefont {M.}~\bibnamefont {Spasenovi{\'{c}}}},
  \bibinfo {author} {\bibfnamefont {J.}~\bibnamefont {Gieseler}}, \bibinfo
  {author} {\bibfnamefont {L.}~\bibnamefont {Rondin}}, \bibinfo {author}
  {\bibfnamefont {L.}~\bibnamefont {Novotny}}, \ and\ \bibinfo {author}
  {\bibfnamefont {R.}~\bibnamefont {Quidant}},\ }\href {\doibase
  10.1038/ncomms15141} {\bibfield  {journal} {\bibinfo  {journal} {Nat.
  Commun.}\ }\textbf {\bibinfo {volume} {8}},\ \bibinfo {pages} {15141}
  (\bibinfo {year} {2017})}\BibitemShut {NoStop}%
\bibitem [{\citenamefont {Rondin}\ \emph {et~al.}(2017)\citenamefont {Rondin},
  \citenamefont {Gieseler}, \citenamefont {Ricci}, \citenamefont {Quidant},
  \citenamefont {Dellago},\ and\ \citenamefont {Novotny}}]{Rondin2017}%
  \BibitemOpen
  \bibfield  {author} {\bibinfo {author} {\bibfnamefont {L.}~\bibnamefont
  {Rondin}}, \bibinfo {author} {\bibfnamefont {J.}~\bibnamefont {Gieseler}},
  \bibinfo {author} {\bibfnamefont {F.}~\bibnamefont {Ricci}}, \bibinfo
  {author} {\bibfnamefont {R.}~\bibnamefont {Quidant}}, \bibinfo {author}
  {\bibfnamefont {C.}~\bibnamefont {Dellago}}, \ and\ \bibinfo {author}
  {\bibfnamefont {L.}~\bibnamefont {Novotny}},\ }\href {\doibase
  10.1038/nnano.2017.198} {\bibfield  {journal} {\bibinfo  {journal} {Nat.
  Nanotechnol.}\ }\textbf {\bibinfo {volume} {12}},\ \bibinfo {pages} {1130}
  (\bibinfo {year} {2017})}\BibitemShut {NoStop}%
\bibitem [{\citenamefont {Gieseler}\ and\ \citenamefont
  {Millen}(2018)}]{Gieseler2018}%
  \BibitemOpen
  \bibfield  {author} {\bibinfo {author} {\bibfnamefont {J.}~\bibnamefont
  {Gieseler}}\ and\ \bibinfo {author} {\bibfnamefont {J.}~\bibnamefont
  {Millen}},\ }\href {\doibase 10.3390/e20050326} {\bibfield  {journal}
  {\bibinfo  {journal} {Entropy}\ }\textbf {\bibinfo {volume} {20}},\ \bibinfo
  {pages} {326} (\bibinfo {year} {2018})}\BibitemShut {NoStop}%
\bibitem [{\citenamefont {Chang}\ \emph {et~al.}(2010)\citenamefont {Chang},
  \citenamefont {Regal}, \citenamefont {Papp}, \citenamefont {Wilson},
  \citenamefont {Ye}, \citenamefont {Painter}, \citenamefont {Kimble},\ and\
  \citenamefont {Zoller}}]{Chang2010}%
  \BibitemOpen
  \bibfield  {author} {\bibinfo {author} {\bibfnamefont {D.~E.}\ \bibnamefont
  {Chang}}, \bibinfo {author} {\bibfnamefont {C.~A.}\ \bibnamefont {Regal}},
  \bibinfo {author} {\bibfnamefont {S.~B.}\ \bibnamefont {Papp}}, \bibinfo
  {author} {\bibfnamefont {D.~J.}\ \bibnamefont {Wilson}}, \bibinfo {author}
  {\bibfnamefont {J.}~\bibnamefont {Ye}}, \bibinfo {author} {\bibfnamefont
  {O.}~\bibnamefont {Painter}}, \bibinfo {author} {\bibfnamefont {H.~J.}\
  \bibnamefont {Kimble}}, \ and\ \bibinfo {author} {\bibfnamefont
  {P.}~\bibnamefont {Zoller}},\ }\href {\doibase 10.1073/pnas.0912969107}
  {\bibfield  {journal} {\bibinfo  {journal} {Proc. Nat. Acad. Sci. USA}\
  }\textbf {\bibinfo {volume} {107}},\ \bibinfo {pages} {1005} (\bibinfo {year}
  {2010})}\BibitemShut {NoStop}%
\bibitem [{\citenamefont {Romero-Isart}\ \emph {et~al.}(2011)\citenamefont
  {Romero-Isart}, \citenamefont {Pflanzer}, \citenamefont {Blaser},
  \citenamefont {Kaltenbaek}, \citenamefont {Kiesel}, \citenamefont
  {Aspelmeyer},\ and\ \citenamefont {Cirac}}]{Romero-Isart2011}%
  \BibitemOpen
  \bibfield  {author} {\bibinfo {author} {\bibfnamefont {O.}~\bibnamefont
  {Romero-Isart}}, \bibinfo {author} {\bibfnamefont {A.~C.}\ \bibnamefont
  {Pflanzer}}, \bibinfo {author} {\bibfnamefont {F.}~\bibnamefont {Blaser}},
  \bibinfo {author} {\bibfnamefont {R.}~\bibnamefont {Kaltenbaek}}, \bibinfo
  {author} {\bibfnamefont {N.}~\bibnamefont {Kiesel}}, \bibinfo {author}
  {\bibfnamefont {M.}~\bibnamefont {Aspelmeyer}}, \ and\ \bibinfo {author}
  {\bibfnamefont {J.~I.}\ \bibnamefont {Cirac}},\ }\href {\doibase
  10.1103/PhysRevLett.107.020405} {\bibfield  {journal} {\bibinfo  {journal}
  {Phys. Rev. Lett.}\ }\textbf {\bibinfo {volume} {107}},\ \bibinfo {pages}
  {020405} (\bibinfo {year} {2011})}\BibitemShut {NoStop}%
\bibitem [{\citenamefont {Gieseler}\ \emph {et~al.}(2012)\citenamefont
  {Gieseler}, \citenamefont {Deutsch}, \citenamefont {Quidant},\ and\
  \citenamefont {Novotny}}]{Gieseler2012}%
  \BibitemOpen
  \bibfield  {author} {\bibinfo {author} {\bibfnamefont {J.}~\bibnamefont
  {Gieseler}}, \bibinfo {author} {\bibfnamefont {B.}~\bibnamefont {Deutsch}},
  \bibinfo {author} {\bibfnamefont {R.}~\bibnamefont {Quidant}}, \ and\
  \bibinfo {author} {\bibfnamefont {L.}~\bibnamefont {Novotny}},\ }\href
  {\doibase 10.1103/PhysRevLett.109.103603} {\bibfield  {journal} {\bibinfo
  {journal} {Phys. Rev. Lett.}\ }\textbf {\bibinfo {volume} {109}},\ \bibinfo
  {pages} {103603} (\bibinfo {year} {2012})}\BibitemShut {NoStop}%
\bibitem [{\citenamefont {Kiesel}\ \emph {et~al.}(2013)\citenamefont {Kiesel},
  \citenamefont {Blaser}, \citenamefont {Deli\'{c}}, \citenamefont {Grass},
  \citenamefont {Kaltenbaek},\ and\ \citenamefont {Aspelmeyer}}]{Kiesel2013}%
  \BibitemOpen
  \bibfield  {author} {\bibinfo {author} {\bibfnamefont {N.}~\bibnamefont
  {Kiesel}}, \bibinfo {author} {\bibfnamefont {F.}~\bibnamefont {Blaser}},
  \bibinfo {author} {\bibfnamefont {U.}~\bibnamefont {Deli\'{c}}}, \bibinfo
  {author} {\bibfnamefont {D.}~\bibnamefont {Grass}}, \bibinfo {author}
  {\bibfnamefont {R.}~\bibnamefont {Kaltenbaek}}, \ and\ \bibinfo {author}
  {\bibfnamefont {M.}~\bibnamefont {Aspelmeyer}},\ }\href {\doibase
  10.1073/pnas.1309167110} {\bibfield  {journal} {\bibinfo  {journal} {Proc.
  Nat. Acad. Sci. USA}\ }\textbf {\bibinfo {volume} {110}},\ \bibinfo {pages}
  {14180} (\bibinfo {year} {2013})}\BibitemShut {NoStop}%
\bibitem [{\citenamefont {Millen}\ \emph {et~al.}(2015)\citenamefont {Millen},
  \citenamefont {Fonseca}, \citenamefont {Mavrogordatos}, \citenamefont
  {Monteiro},\ and\ \citenamefont {Barker}}]{Millen2015}%
  \BibitemOpen
  \bibfield  {author} {\bibinfo {author} {\bibfnamefont {J.}~\bibnamefont
  {Millen}}, \bibinfo {author} {\bibfnamefont {P.~Z.~G.}\ \bibnamefont
  {Fonseca}}, \bibinfo {author} {\bibfnamefont {T.}~\bibnamefont
  {Mavrogordatos}}, \bibinfo {author} {\bibfnamefont {T.~S.}\ \bibnamefont
  {Monteiro}}, \ and\ \bibinfo {author} {\bibfnamefont {P.~F.}\ \bibnamefont
  {Barker}},\ }\href {\doibase 10.1103/PhysRevLett.114.123602} {\bibfield
  {journal} {\bibinfo  {journal} {Phys. Rev. Lett.}\ }\textbf {\bibinfo
  {volume} {114}},\ \bibinfo {pages} {123602} (\bibinfo {year}
  {2015})}\BibitemShut {NoStop}%
\bibitem [{\citenamefont {Jain}\ \emph {et~al.}(2016)\citenamefont {Jain},
  \citenamefont {Gieseler}, \citenamefont {Moritz}, \citenamefont {Dellago},
  \citenamefont {Quidant},\ and\ \citenamefont {Novotny}}]{Jain2016}%
  \BibitemOpen
  \bibfield  {author} {\bibinfo {author} {\bibfnamefont {V.}~\bibnamefont
  {Jain}}, \bibinfo {author} {\bibfnamefont {J.}~\bibnamefont {Gieseler}},
  \bibinfo {author} {\bibfnamefont {C.}~\bibnamefont {Moritz}}, \bibinfo
  {author} {\bibfnamefont {C.}~\bibnamefont {Dellago}}, \bibinfo {author}
  {\bibfnamefont {R.}~\bibnamefont {Quidant}}, \ and\ \bibinfo {author}
  {\bibfnamefont {L.}~\bibnamefont {Novotny}},\ }\href {\doibase
  10.1103/PhysRevLett.116.243601} {\bibfield  {journal} {\bibinfo  {journal}
  {Phys. Rev. Lett.}\ }\textbf {\bibinfo {volume} {116}},\ \bibinfo {pages}
  {243601} (\bibinfo {year} {2016})}\BibitemShut {NoStop}%
\bibitem [{\citenamefont {Vovrosh}\ \emph {et~al.}(2017)\citenamefont
  {Vovrosh}, \citenamefont {Rashid}, \citenamefont {Hempston}, \citenamefont
  {Bateman}, \citenamefont {Paternostro},\ and\ \citenamefont
  {Ulbricht}}]{Vovrosh2017}%
  \BibitemOpen
  \bibfield  {author} {\bibinfo {author} {\bibfnamefont {J.}~\bibnamefont
  {Vovrosh}}, \bibinfo {author} {\bibfnamefont {M.}~\bibnamefont {Rashid}},
  \bibinfo {author} {\bibfnamefont {D.}~\bibnamefont {Hempston}}, \bibinfo
  {author} {\bibfnamefont {J.}~\bibnamefont {Bateman}}, \bibinfo {author}
  {\bibfnamefont {M.}~\bibnamefont {Paternostro}}, \ and\ \bibinfo {author}
  {\bibfnamefont {H.}~\bibnamefont {Ulbricht}},\ }\href {\doibase
  10.1364/JOSAB.34.001421} {\bibfield  {journal} {\bibinfo  {journal} {J. Opt.
  Soc. Am. B}\ }\textbf {\bibinfo {volume} {34}},\ \bibinfo {pages} {1421}
  (\bibinfo {year} {2017})}\BibitemShut {NoStop}%
\bibitem [{\citenamefont {Monteiro}\ \emph {et~al.}(2017)\citenamefont
  {Monteiro}, \citenamefont {Ghosh}, \citenamefont {Fine},\ and\ \citenamefont
  {Moore}}]{Monteiro2017}%
  \BibitemOpen
  \bibfield  {author} {\bibinfo {author} {\bibfnamefont {F.}~\bibnamefont
  {Monteiro}}, \bibinfo {author} {\bibfnamefont {S.}~\bibnamefont {Ghosh}},
  \bibinfo {author} {\bibfnamefont {A.~G.}\ \bibnamefont {Fine}}, \ and\
  \bibinfo {author} {\bibfnamefont {D.~C.}\ \bibnamefont {Moore}},\ }\href
  {\doibase 10.1103/PhysRevA.96.063841} {\bibfield  {journal} {\bibinfo
  {journal} {Phys. Rev. A}\ }\textbf {\bibinfo {volume} {96}},\ \bibinfo
  {pages} {063841} (\bibinfo {year} {2017})}\BibitemShut {NoStop}%
\bibitem [{\citenamefont {Windey}\ \emph {et~al.}(2019)\citenamefont {Windey},
  \citenamefont {Gonzalez-Ballestero}, \citenamefont {Maurer}, \citenamefont
  {Novotny}, \citenamefont {Romero-Isart},\ and\ \citenamefont
  {Reimann}}]{Windey2019}%
  \BibitemOpen
  \bibfield  {author} {\bibinfo {author} {\bibfnamefont {D.}~\bibnamefont
  {Windey}}, \bibinfo {author} {\bibfnamefont {C.}~\bibnamefont
  {Gonzalez-Ballestero}}, \bibinfo {author} {\bibfnamefont {P.}~\bibnamefont
  {Maurer}}, \bibinfo {author} {\bibfnamefont {L.}~\bibnamefont {Novotny}},
  \bibinfo {author} {\bibfnamefont {O.}~\bibnamefont {Romero-Isart}}, \ and\
  \bibinfo {author} {\bibfnamefont {R.}~\bibnamefont {Reimann}},\ }\href
  {\doibase 10.1103/PhysRevLett.122.123601} {\bibfield  {journal} {\bibinfo
  {journal} {Phys. Rev. Lett.}\ }\textbf {\bibinfo {volume} {122}},\ \bibinfo
  {pages} {123601} (\bibinfo {year} {2019})}\BibitemShut {NoStop}%
\bibitem [{\citenamefont {Deli{\'{c}}}\ \emph {et~al.}(2019)\citenamefont
  {Deli{\'{c}}}, \citenamefont {Reisenbauer}, \citenamefont {Grass},
  \citenamefont {Kiesel}, \citenamefont {Vuleti{\'{c}}},\ and\ \citenamefont
  {Aspelmeyer}}]{Delic2019}%
  \BibitemOpen
  \bibfield  {author} {\bibinfo {author} {\bibfnamefont {U.}~\bibnamefont
  {Deli{\'{c}}}}, \bibinfo {author} {\bibfnamefont {M.}~\bibnamefont
  {Reisenbauer}}, \bibinfo {author} {\bibfnamefont {D.}~\bibnamefont {Grass}},
  \bibinfo {author} {\bibfnamefont {N.}~\bibnamefont {Kiesel}}, \bibinfo
  {author} {\bibfnamefont {V.}~\bibnamefont {Vuleti{\'{c}}}}, \ and\ \bibinfo
  {author} {\bibfnamefont {M.}~\bibnamefont {Aspelmeyer}},\ }\href {\doibase
  10.1103/PhysRevLett.122.123602} {\bibfield  {journal} {\bibinfo  {journal}
  {Phys. Rev. Lett.}\ }\textbf {\bibinfo {volume} {122}},\ \bibinfo {pages}
  {123602} (\bibinfo {year} {2019})}\BibitemShut {NoStop}%
\bibitem [{\citenamefont {Tebbenjohanns}\ \emph {et~al.}(2020)\citenamefont
  {Tebbenjohanns}, \citenamefont {Frimmer}, \citenamefont {Jain}, \citenamefont
  {Windey},\ and\ \citenamefont {Novotny}}]{Tebbenjohanns2020}%
  \BibitemOpen
  \bibfield  {author} {\bibinfo {author} {\bibfnamefont {F.}~\bibnamefont
  {Tebbenjohanns}}, \bibinfo {author} {\bibfnamefont {M.}~\bibnamefont
  {Frimmer}}, \bibinfo {author} {\bibfnamefont {V.}~\bibnamefont {Jain}},
  \bibinfo {author} {\bibfnamefont {D.}~\bibnamefont {Windey}}, \ and\ \bibinfo
  {author} {\bibfnamefont {L.}~\bibnamefont {Novotny}},\ }\href {\doibase
  10.1103/PhysRevLett.124.013603} {\bibfield  {journal} {\bibinfo  {journal}
  {Phys. Rev. Lett.}\ }\textbf {\bibinfo {volume} {124}},\ \bibinfo {pages}
  {013603} (\bibinfo {year} {2020})}\BibitemShut {NoStop}%
\bibitem [{\citenamefont {Arita}\ \emph {et~al.}(2013)\citenamefont {Arita},
  \citenamefont {Mazilu},\ and\ \citenamefont {Dholakia}}]{Arita2013}%
  \BibitemOpen
  \bibfield  {author} {\bibinfo {author} {\bibfnamefont {Y.}~\bibnamefont
  {Arita}}, \bibinfo {author} {\bibfnamefont {M.}~\bibnamefont {Mazilu}}, \
  and\ \bibinfo {author} {\bibfnamefont {K.}~\bibnamefont {Dholakia}},\ }\href
  {\doibase 10.1038/ncomms3374} {\bibfield  {journal} {\bibinfo  {journal}
  {Nat. Commun.}\ }\textbf {\bibinfo {volume} {4}},\ \bibinfo {pages} {2374}
  (\bibinfo {year} {2013})}\BibitemShut {NoStop}%
\bibitem [{\citenamefont {Arita}\ \emph {et~al.}(2015)\citenamefont {Arita},
  \citenamefont {Mazilu}, \citenamefont {Vettenburg}, \citenamefont {Wright},\
  and\ \citenamefont {Dholakia}}]{Arita2015}%
  \BibitemOpen
  \bibfield  {author} {\bibinfo {author} {\bibfnamefont {Y.}~\bibnamefont
  {Arita}}, \bibinfo {author} {\bibfnamefont {M.}~\bibnamefont {Mazilu}},
  \bibinfo {author} {\bibfnamefont {T.}~\bibnamefont {Vettenburg}}, \bibinfo
  {author} {\bibfnamefont {E.~M.}\ \bibnamefont {Wright}}, \ and\ \bibinfo
  {author} {\bibfnamefont {K.}~\bibnamefont {Dholakia}},\ }\href {\doibase
  10.1364/ol.40.004751} {\bibfield  {journal} {\bibinfo  {journal} {Opt.
  Lett.}\ }\textbf {\bibinfo {volume} {40}},\ \bibinfo {pages} {4751} (\bibinfo
  {year} {2015})}\BibitemShut {NoStop}%
\bibitem [{\citenamefont {Hoang}\ \emph {et~al.}(2016)\citenamefont {Hoang},
  \citenamefont {Ma}, \citenamefont {Ahn}, \citenamefont {Bang}, \citenamefont
  {Robicheaux}, \citenamefont {Yin},\ and\ \citenamefont {Li}}]{Hoang2016}%
  \BibitemOpen
  \bibfield  {author} {\bibinfo {author} {\bibfnamefont {T.~M.}\ \bibnamefont
  {Hoang}}, \bibinfo {author} {\bibfnamefont {Y.}~\bibnamefont {Ma}}, \bibinfo
  {author} {\bibfnamefont {J.}~\bibnamefont {Ahn}}, \bibinfo {author}
  {\bibfnamefont {J.}~\bibnamefont {Bang}}, \bibinfo {author} {\bibfnamefont
  {F.}~\bibnamefont {Robicheaux}}, \bibinfo {author} {\bibfnamefont {Z.~Q.}\
  \bibnamefont {Yin}}, \ and\ \bibinfo {author} {\bibfnamefont
  {T.}~\bibnamefont {Li}},\ }\href {\doibase 10.1103/PhysRevLett.117.123604}
  {\bibfield  {journal} {\bibinfo  {journal} {Phys. Rev. Lett.}\ }\textbf
  {\bibinfo {volume} {117}},\ \bibinfo {pages} {1} (\bibinfo {year}
  {2016})}\BibitemShut {NoStop}%
\bibitem [{\citenamefont {Kuhn}\ \emph {et~al.}(2015)\citenamefont {Kuhn},
  \citenamefont {Asenbaum}, \citenamefont {Kosloff}, \citenamefont {Sclafani},
  \citenamefont {Stickler}, \citenamefont {Nimmrichter}, \citenamefont
  {Hornberger}, \citenamefont {Cheshnovsky}, \citenamefont {Patolsky},\ and\
  \citenamefont {Arndt}}]{Kuhn2015}%
  \BibitemOpen
  \bibfield  {author} {\bibinfo {author} {\bibfnamefont {S.}~\bibnamefont
  {Kuhn}}, \bibinfo {author} {\bibfnamefont {P.}~\bibnamefont {Asenbaum}},
  \bibinfo {author} {\bibfnamefont {A.}~\bibnamefont {Kosloff}}, \bibinfo
  {author} {\bibfnamefont {M.}~\bibnamefont {Sclafani}}, \bibinfo {author}
  {\bibfnamefont {B.~A.}\ \bibnamefont {Stickler}}, \bibinfo {author}
  {\bibfnamefont {S.}~\bibnamefont {Nimmrichter}}, \bibinfo {author}
  {\bibfnamefont {K.}~\bibnamefont {Hornberger}}, \bibinfo {author}
  {\bibfnamefont {O.}~\bibnamefont {Cheshnovsky}}, \bibinfo {author}
  {\bibfnamefont {F.}~\bibnamefont {Patolsky}}, \ and\ \bibinfo {author}
  {\bibfnamefont {M.}~\bibnamefont {Arndt}},\ }\href {\doibase
  10.1021/acs.nanolett.5b02302} {\bibfield  {journal} {\bibinfo  {journal}
  {Nano Lett.}\ }\textbf {\bibinfo {volume} {15}},\ \bibinfo {pages} {5604}
  (\bibinfo {year} {2015})}\BibitemShut {NoStop}%
\bibitem [{\citenamefont {Kuhn}\ \emph
  {et~al.}(2017{\natexlab{a}})\citenamefont {Kuhn}, \citenamefont {Kosloff},
  \citenamefont {Stickler}, \citenamefont {Patolsky}, \citenamefont
  {Hornberger}, \citenamefont {Arndt},\ and\ \citenamefont
  {Millen}}]{Kuhn2017control}%
  \BibitemOpen
  \bibfield  {author} {\bibinfo {author} {\bibfnamefont {S.}~\bibnamefont
  {Kuhn}}, \bibinfo {author} {\bibfnamefont {A.}~\bibnamefont {Kosloff}},
  \bibinfo {author} {\bibfnamefont {B.~A.}\ \bibnamefont {Stickler}}, \bibinfo
  {author} {\bibfnamefont {F.}~\bibnamefont {Patolsky}}, \bibinfo {author}
  {\bibfnamefont {K.}~\bibnamefont {Hornberger}}, \bibinfo {author}
  {\bibfnamefont {M.}~\bibnamefont {Arndt}}, \ and\ \bibinfo {author}
  {\bibfnamefont {J.}~\bibnamefont {Millen}},\ }\href {\doibase
  10.1364/OPTICA.4.000356} {\bibfield  {journal} {\bibinfo  {journal} {Optica}\
  }\textbf {\bibinfo {volume} {4}},\ \bibinfo {pages} {356} (\bibinfo {year}
  {2017}{\natexlab{a}})}\BibitemShut {NoStop}%
\bibitem [{\citenamefont {Kuhn}\ \emph
  {et~al.}(2017{\natexlab{b}})\citenamefont {Kuhn}, \citenamefont {Stickler},
  \citenamefont {Kosloff}, \citenamefont {Patolsky}, \citenamefont
  {Hornberger}, \citenamefont {Arndt},\ and\ \citenamefont
  {Millen}}]{Kuhn2017stable}%
  \BibitemOpen
  \bibfield  {author} {\bibinfo {author} {\bibfnamefont {S.}~\bibnamefont
  {Kuhn}}, \bibinfo {author} {\bibfnamefont {B.~A.}\ \bibnamefont {Stickler}},
  \bibinfo {author} {\bibfnamefont {A.}~\bibnamefont {Kosloff}}, \bibinfo
  {author} {\bibfnamefont {F.}~\bibnamefont {Patolsky}}, \bibinfo {author}
  {\bibfnamefont {K.}~\bibnamefont {Hornberger}}, \bibinfo {author}
  {\bibfnamefont {M.}~\bibnamefont {Arndt}}, \ and\ \bibinfo {author}
  {\bibfnamefont {J.}~\bibnamefont {Millen}},\ }\href {\doibase
  10.1038/s41467-017-01902-9} {\bibfield  {journal} {\bibinfo  {journal} {Nat.
  Commun.}\ }\textbf {\bibinfo {volume} {8}},\ \bibinfo {pages} {1} (\bibinfo
  {year} {2017}{\natexlab{b}})}\BibitemShut {NoStop}%
\bibitem [{\citenamefont {Reimann}\ \emph {et~al.}(2018)\citenamefont
  {Reimann}, \citenamefont {Doderer}, \citenamefont {Hebestreit}, \citenamefont
  {Diehl}, \citenamefont {Frimmer}, \citenamefont {Windey}, \citenamefont
  {Tebbenjohanns},\ and\ \citenamefont {Novotny}}]{Reimann2018}%
  \BibitemOpen
  \bibfield  {author} {\bibinfo {author} {\bibfnamefont {R.}~\bibnamefont
  {Reimann}}, \bibinfo {author} {\bibfnamefont {M.}~\bibnamefont {Doderer}},
  \bibinfo {author} {\bibfnamefont {E.}~\bibnamefont {Hebestreit}}, \bibinfo
  {author} {\bibfnamefont {R.}~\bibnamefont {Diehl}}, \bibinfo {author}
  {\bibfnamefont {M.}~\bibnamefont {Frimmer}}, \bibinfo {author} {\bibfnamefont
  {D.}~\bibnamefont {Windey}}, \bibinfo {author} {\bibfnamefont
  {F.}~\bibnamefont {Tebbenjohanns}}, \ and\ \bibinfo {author} {\bibfnamefont
  {L.}~\bibnamefont {Novotny}},\ }\href {\doibase
  10.1103/PhysRevLett.121.033602} {\bibfield  {journal} {\bibinfo  {journal}
  {Phys. Rev. Lett.}\ }\textbf {\bibinfo {volume} {121}},\ \bibinfo {pages} {1}
  (\bibinfo {year} {2018})}\BibitemShut {NoStop}%
\bibitem [{\citenamefont {Ahn}\ \emph {et~al.}(2018)\citenamefont {Ahn},
  \citenamefont {Xu}, \citenamefont {Bang}, \citenamefont {Deng}, \citenamefont
  {Hoang}, \citenamefont {Han}, \citenamefont {Ma},\ and\ \citenamefont
  {Li}}]{Ahn2018}%
  \BibitemOpen
  \bibfield  {author} {\bibinfo {author} {\bibfnamefont {J.}~\bibnamefont
  {Ahn}}, \bibinfo {author} {\bibfnamefont {Z.}~\bibnamefont {Xu}}, \bibinfo
  {author} {\bibfnamefont {J.}~\bibnamefont {Bang}}, \bibinfo {author}
  {\bibfnamefont {Y.~H.}\ \bibnamefont {Deng}}, \bibinfo {author}
  {\bibfnamefont {T.~M.}\ \bibnamefont {Hoang}}, \bibinfo {author}
  {\bibfnamefont {Q.}~\bibnamefont {Han}}, \bibinfo {author} {\bibfnamefont
  {R.~M.}\ \bibnamefont {Ma}}, \ and\ \bibinfo {author} {\bibfnamefont
  {T.}~\bibnamefont {Li}},\ }\href {\doibase 10.1103/PhysRevLett.121.033603}
  {\bibfield  {journal} {\bibinfo  {journal} {Phys. Rev. Lett.}\ }\textbf
  {\bibinfo {volume} {121}},\ \bibinfo {pages} {33603} (\bibinfo {year}
  {2018})}\BibitemShut {NoStop}%
\bibitem [{\citenamefont {Monteiro}\ \emph {et~al.}(2018)\citenamefont
  {Monteiro}, \citenamefont {Ghosh}, \citenamefont {{Van Assendelft}},\ and\
  \citenamefont {Moore}}]{Monteiro2018}%
  \BibitemOpen
  \bibfield  {author} {\bibinfo {author} {\bibfnamefont {F.}~\bibnamefont
  {Monteiro}}, \bibinfo {author} {\bibfnamefont {S.}~\bibnamefont {Ghosh}},
  \bibinfo {author} {\bibfnamefont {E.~C.}\ \bibnamefont {{Van Assendelft}}}, \
  and\ \bibinfo {author} {\bibfnamefont {D.~C.}\ \bibnamefont {Moore}},\ }\href
  {\doibase 10.1103/PhysRevA.97.051802} {\bibfield  {journal} {\bibinfo
  {journal} {Phys. Rev. A}\ }\textbf {\bibinfo {volume} {97}},\ \bibinfo
  {pages} {1} (\bibinfo {year} {2018})}\BibitemShut {NoStop}%
\bibitem [{\citenamefont {Rider}\ \emph {et~al.}(2019)\citenamefont {Rider},
  \citenamefont {Blakemore}, \citenamefont {Kawasaki}, \citenamefont {Priel},
  \citenamefont {Roy},\ and\ \citenamefont {Gratta}}]{Rider2019}%
  \BibitemOpen
  \bibfield  {author} {\bibinfo {author} {\bibfnamefont {A.~D.}\ \bibnamefont
  {Rider}}, \bibinfo {author} {\bibfnamefont {C.~P.}\ \bibnamefont
  {Blakemore}}, \bibinfo {author} {\bibfnamefont {A.}~\bibnamefont {Kawasaki}},
  \bibinfo {author} {\bibfnamefont {N.}~\bibnamefont {Priel}}, \bibinfo
  {author} {\bibfnamefont {S.}~\bibnamefont {Roy}}, \ and\ \bibinfo {author}
  {\bibfnamefont {G.}~\bibnamefont {Gratta}},\ }\href {\doibase
  10.1103/PhysRevA.99.041802} {\bibfield  {journal} {\bibinfo  {journal} {Phys.
  Rev. A}\ }\textbf {\bibinfo {volume} {99}},\ \bibinfo {pages} {041802}
  (\bibinfo {year} {2019})}\BibitemShut {NoStop}%
\bibitem [{\citenamefont {Delord}\ \emph {et~al.}(2019)\citenamefont {Delord},
  \citenamefont {Huillery}, \citenamefont {Nicolas},\ and\ \citenamefont
  {H{\'{e}}tet}}]{Delord2019}%
  \BibitemOpen
  \bibfield  {author} {\bibinfo {author} {\bibfnamefont {T.}~\bibnamefont
  {Delord}}, \bibinfo {author} {\bibfnamefont {P.}~\bibnamefont {Huillery}},
  \bibinfo {author} {\bibfnamefont {L.}~\bibnamefont {Nicolas}}, \ and\
  \bibinfo {author} {\bibfnamefont {G.}~\bibnamefont {H{\'{e}}tet}},\ }\href
  {\doibase 10.1038/s41586-020-2133-z} {\bibfield  {journal} {\bibinfo
  {journal} {Nature (London)}\ }\textbf {\bibinfo {volume} {580}} (\bibinfo
  {year} {2019}),\ 10.1038/s41586-020-2133-z}\BibitemShut {NoStop}%
\bibitem [{\citenamefont {Ahn}\ \emph {et~al.}(2020)\citenamefont {Ahn},
  \citenamefont {Xu}, \citenamefont {Bang}, \citenamefont {Ju}, \citenamefont
  {Gao},\ and\ \citenamefont {Li}}]{Ahn2020}%
  \BibitemOpen
  \bibfield  {author} {\bibinfo {author} {\bibfnamefont {J.}~\bibnamefont
  {Ahn}}, \bibinfo {author} {\bibfnamefont {Z.}~\bibnamefont {Xu}}, \bibinfo
  {author} {\bibfnamefont {J.}~\bibnamefont {Bang}}, \bibinfo {author}
  {\bibfnamefont {P.}~\bibnamefont {Ju}}, \bibinfo {author} {\bibfnamefont
  {X.}~\bibnamefont {Gao}}, \ and\ \bibinfo {author} {\bibfnamefont
  {T.}~\bibnamefont {Li}},\ }\href {\doibase 10.1038/s41565-019-0605-9}
  {\bibfield  {journal} {\bibinfo  {journal} {Nature Nanotechnology}\ }\textbf
  {\bibinfo {volume} {15}},\ \bibinfo {pages} {89} (\bibinfo {year}
  {2020})}\BibitemShut {NoStop}%
\bibitem [{\citenamefont {Shi}\ and\ \citenamefont
  {Bhattacharya}(2016)}]{Shi2016}%
  \BibitemOpen
  \bibfield  {author} {\bibinfo {author} {\bibfnamefont {H.}~\bibnamefont
  {Shi}}\ and\ \bibinfo {author} {\bibfnamefont {M.}~\bibnamefont
  {Bhattacharya}},\ }\href {\doibase 10.1088/0953-4075/49/15/153001} {\bibfield
   {journal} {\bibinfo  {journal} {J. Phys. B}\ }\textbf {\bibinfo {volume}
  {49}},\ \bibinfo {pages} {153001} (\bibinfo {year} {2016})}\BibitemShut
  {NoStop}%
\bibitem [{\citenamefont {Stickler}\ \emph
  {et~al.}(2018{\natexlab{a}})\citenamefont {Stickler}, \citenamefont
  {Schrinski},\ and\ \citenamefont {Hornberger}}]{Stickler2018}%
  \BibitemOpen
  \bibfield  {author} {\bibinfo {author} {\bibfnamefont {B.~A.}\ \bibnamefont
  {Stickler}}, \bibinfo {author} {\bibfnamefont {B.}~\bibnamefont {Schrinski}},
  \ and\ \bibinfo {author} {\bibfnamefont {K.}~\bibnamefont {Hornberger}},\
  }\href {\doibase 10.1103/PhysRevLett.121.040401} {\bibfield  {journal}
  {\bibinfo  {journal} {Phys. Rev. Lett.}\ }\textbf {\bibinfo {volume} {121}},\
  \bibinfo {pages} {40401} (\bibinfo {year} {2018}{\natexlab{a}})}\BibitemShut
  {NoStop}%
\bibitem [{\citenamefont {Stickler}\ \emph
  {et~al.}(2018{\natexlab{b}})\citenamefont {Stickler}, \citenamefont
  {Papendell}, \citenamefont {Kuhn}, \citenamefont {Schrinski}, \citenamefont
  {Millen}, \citenamefont {Arndt},\ and\ \citenamefont
  {Hornberger}}]{Stickler2018rotationquantum}%
  \BibitemOpen
  \bibfield  {author} {\bibinfo {author} {\bibfnamefont {B.~A.}\ \bibnamefont
  {Stickler}}, \bibinfo {author} {\bibfnamefont {B.}~\bibnamefont {Papendell}},
  \bibinfo {author} {\bibfnamefont {S.}~\bibnamefont {Kuhn}}, \bibinfo {author}
  {\bibfnamefont {B.}~\bibnamefont {Schrinski}}, \bibinfo {author}
  {\bibfnamefont {J.}~\bibnamefont {Millen}}, \bibinfo {author} {\bibfnamefont
  {M.}~\bibnamefont {Arndt}}, \ and\ \bibinfo {author} {\bibfnamefont
  {K.}~\bibnamefont {Hornberger}},\ }\href {\doibase 10.1088/1367-2630/aaece4}
  {\bibfield  {journal} {\bibinfo  {journal} {New J. Phys.}\ }\textbf {\bibinfo
  {volume} {20}},\ \bibinfo {pages} {122001} (\bibinfo {year}
  {2018}{\natexlab{b}})}\BibitemShut {NoStop}%
\bibitem [{\citenamefont {Blakemore}\ \emph {et~al.}(2020)\citenamefont
  {Blakemore}, \citenamefont {Martin}, \citenamefont {Fieguth}, \citenamefont
  {Kawasaki}, \citenamefont {Priel}, \citenamefont {Rider},\ and\ \citenamefont
  {Gratta}}]{Blakemore2020}%
  \BibitemOpen
  \bibfield  {author} {\bibinfo {author} {\bibfnamefont {C.~P.}\ \bibnamefont
  {Blakemore}}, \bibinfo {author} {\bibfnamefont {D.}~\bibnamefont {Martin}},
  \bibinfo {author} {\bibfnamefont {A.}~\bibnamefont {Fieguth}}, \bibinfo
  {author} {\bibfnamefont {A.}~\bibnamefont {Kawasaki}}, \bibinfo {author}
  {\bibfnamefont {N.}~\bibnamefont {Priel}}, \bibinfo {author} {\bibfnamefont
  {A.~D.}\ \bibnamefont {Rider}}, \ and\ \bibinfo {author} {\bibfnamefont
  {G.}~\bibnamefont {Gratta}},\ }\href {\doibase 10.1116/1.5139638} {\bibfield
  {journal} {\bibinfo  {journal} {J. Vac. Sci. Technol. B}\ }\textbf {\bibinfo
  {volume} {38}},\ \bibinfo {pages} {024201} (\bibinfo {year}
  {2020})}\BibitemShut {NoStop}%
\bibitem [{\citenamefont {Manjavacas}\ and\ \citenamefont {{Garc{\'{i}}a De
  Abajo}}(2010)}]{Manjavacas2010}%
  \BibitemOpen
  \bibfield  {author} {\bibinfo {author} {\bibfnamefont {A.}~\bibnamefont
  {Manjavacas}}\ and\ \bibinfo {author} {\bibfnamefont {F.~J.}\ \bibnamefont
  {{Garc{\'{i}}a De Abajo}}},\ }\href {\doibase 10.1103/PhysRevLett.105.113601}
  {\bibfield  {journal} {\bibinfo  {journal} {Phys. Rev. Lett.}\ }\textbf
  {\bibinfo {volume} {105}},\ \bibinfo {pages} {1} (\bibinfo {year}
  {2010})}\BibitemShut {NoStop}%
\bibitem [{\citenamefont {Zhao}\ \emph {et~al.}(2012)\citenamefont {Zhao},
  \citenamefont {Manjavacas}, \citenamefont {{Garc{\'{i}}a De Abajo}},\ and\
  \citenamefont {Pendry}}]{Zhao2012}%
  \BibitemOpen
  \bibfield  {author} {\bibinfo {author} {\bibfnamefont {R.}~\bibnamefont
  {Zhao}}, \bibinfo {author} {\bibfnamefont {A.}~\bibnamefont {Manjavacas}},
  \bibinfo {author} {\bibfnamefont {F.~J.}\ \bibnamefont {{Garc{\'{i}}a De
  Abajo}}}, \ and\ \bibinfo {author} {\bibfnamefont {J.~B.}\ \bibnamefont
  {Pendry}},\ }\href {\doibase 10.1103/PhysRevLett.109.123604} {\bibfield
  {journal} {\bibinfo  {journal} {Phys. Rev. Lett.}\ }\textbf {\bibinfo
  {volume} {109}},\ \bibinfo {pages} {1} (\bibinfo {year} {2012})}\BibitemShut
  {NoStop}%
\bibitem [{\citenamefont {Manjavacas}\ \emph {et~al.}(2017)\citenamefont
  {Manjavacas}, \citenamefont {Rodr{\'{i}}guez-Fortu{\~{n}}o}, \citenamefont
  {{Javier Garc{\'{i}}a De Abajo}},\ and\ \citenamefont
  {Zayats}}]{Manjavacas2017}%
  \BibitemOpen
  \bibfield  {author} {\bibinfo {author} {\bibfnamefont {A.}~\bibnamefont
  {Manjavacas}}, \bibinfo {author} {\bibfnamefont {F.~J.}\ \bibnamefont
  {Rodr{\'{i}}guez-Fortu{\~{n}}o}}, \bibinfo {author} {\bibfnamefont
  {F.}~\bibnamefont {{Javier Garc{\'{i}}a De Abajo}}}, \ and\ \bibinfo {author}
  {\bibfnamefont {A.~V.}\ \bibnamefont {Zayats}},\ }\href {\doibase
  10.1103/PhysRevLett.118.133605} {\bibfield  {journal} {\bibinfo  {journal}
  {Phys. Rev. Lett.}\ }\textbf {\bibinfo {volume} {118}},\ \bibinfo {pages} {1}
  (\bibinfo {year} {2017})}\BibitemShut {NoStop}%
\bibitem [{\citenamefont {Xu}\ and\ \citenamefont {Li}(2017)}]{Xu2017}%
  \BibitemOpen
  \bibfield  {author} {\bibinfo {author} {\bibfnamefont {Z.}~\bibnamefont
  {Xu}}\ and\ \bibinfo {author} {\bibfnamefont {T.}~\bibnamefont {Li}},\ }\href
  {\doibase 10.1103/PhysRevA.96.033843} {\bibfield  {journal} {\bibinfo
  {journal} {Phys. Rev. A}\ }\textbf {\bibinfo {volume} {96}},\ \bibinfo
  {pages} {1} (\bibinfo {year} {2017})}\BibitemShut {NoStop}%
\bibitem [{\citenamefont {Hebestreit}\ \emph {et~al.}(2018)\citenamefont
  {Hebestreit}, \citenamefont {Frimmer}, \citenamefont {Reimann}, \citenamefont
  {Dellago}, \citenamefont {Ricci},\ and\ \citenamefont
  {Novotny}}]{Hebestreit2018Calibration}%
  \BibitemOpen
  \bibfield  {author} {\bibinfo {author} {\bibfnamefont {E.}~\bibnamefont
  {Hebestreit}}, \bibinfo {author} {\bibfnamefont {M.}~\bibnamefont {Frimmer}},
  \bibinfo {author} {\bibfnamefont {R.}~\bibnamefont {Reimann}}, \bibinfo
  {author} {\bibfnamefont {C.}~\bibnamefont {Dellago}}, \bibinfo {author}
  {\bibfnamefont {F.}~\bibnamefont {Ricci}}, \ and\ \bibinfo {author}
  {\bibfnamefont {L.}~\bibnamefont {Novotny}},\ }\href {\doibase
  10.1063/1.5017119} {\bibfield  {journal} {\bibinfo  {journal} {Rev. Sci.
  Instrum.}\ }\textbf {\bibinfo {volume} {89}},\ \bibinfo {pages} {033111}
  (\bibinfo {year} {2018})}\BibitemShut {NoStop}%
\bibitem [{\citenamefont {Gieseler}(2014)}]{GieselerThesis}%
  \BibitemOpen
  \bibfield  {author} {\bibinfo {author} {\bibfnamefont {J.}~\bibnamefont
  {Gieseler}},\ }\href@noop {} {Ph.D. thesis},\ \bibinfo  {school} {Universitat
  Politecnica de Catalunya} (\bibinfo {year} {2014})\BibitemShut {NoStop}%
\bibitem [{\citenamefont {Kubo}(1966)}]{kubo1966fluctuation}%
  \BibitemOpen
  \bibfield  {author} {\bibinfo {author} {\bibfnamefont {R.}~\bibnamefont
  {Kubo}},\ }\href {\doibase 10.1088/0034-4885/29/1/306} {\bibfield  {journal}
  {\bibinfo  {journal} {Rep. Prog. Phys.}\ }\textbf {\bibinfo {volume} {29}},\
  \bibinfo {pages} {255} (\bibinfo {year} {1966})}\BibitemShut {NoStop}%
\bibitem [{\citenamefont {Taylor}(2005)}]{Taylorbook}%
  \BibitemOpen
  \bibfield  {author} {\bibinfo {author} {\bibfnamefont {J.~R.}\ \bibnamefont
  {Taylor}},\ }\href@noop {} {\emph {\bibinfo {title} {Classical Mechanics}}}\
  (\bibinfo  {publisher} {University Science Books},\ \bibinfo {address} {Mill
  Valley, CA,},\ \bibinfo {year} {2005})\ p.\ \bibinfo {pages}
  {411}\BibitemShut {NoStop}%
\bibitem [{\citenamefont {Ricci}(2019)}]{RicciThesis}%
  \BibitemOpen
  \bibfield  {author} {\bibinfo {author} {\bibfnamefont {F.}~\bibnamefont
  {Ricci}},\ }\href@noop {} {Ph.D. thesis},\ \bibinfo  {school} {Universitat
  Politecnica de Catalunya} (\bibinfo {year} {2019})\BibitemShut {NoStop}%
\end{thebibliography}%

\end{document}